\documentclass{aa}  
\usepackage{graphics}
\begin{document}  
   \thesaurus{06    
              (08.02.2;   
               08.06.3;
               08.05.1;
               08.09.2 HV\,2543)}  
   \title{Determination of masses and radii of the massive eclipsing binary 
system \object{HV\,2543} in the Large Magellanic Cloud  
}

\author{  
P.G. Ostrov \inst{1}\fnmsep\thanks{Visiting Astronomer, Complejo   
Astron\'omico El Leoncito operated under agreement between the Consejo Nacional
de Investigaciones Cient\'{\i}ficas y T\'ecnicas de la Rep\'ublica Argentina   
and the National Universities of La Plata, C\'ordoba and San Juan.}, 
E. Lapasset \inst{2,3}
\and
N.I. Morrell \inst{1,3}  
}  
  
\offprints{P.G. Ostrov}
\mail{ostrov@fcaglp.edu.ar}
  
\institute{  
Facultad de Ciencias Astron\'omicas y Geof\'{\i}sicas, Universidad   
Nacional de La Plata, Paseo del Bosque S/N, 1900 La Plata, Argentina 
\and  
Observatorio Astron\'omico, Universidad Nacional de C\'ordoba, Laprida 854,   
5000 C\'ordoba, Argentina  
\and  
Member of the Carrera del Investigador Cient\'{\i}fico, CONICET, Argentina.  
}  
  
\date{Received September 15, 1996; accepted March 16, 1997}

\titlerunning{The eclipsing binary \object{HV\,2543}}  
\authorrunning{Ostrov et al.}
  
   \maketitle  
  
   \begin{abstract}  
  
 We present a $V$ light curve of the eclipsing binary system   
\object{HV\,2543} (\object{Sk--67${\degr }$117}) in the Large Magellanic Cloud 
based on CCD images acquired between the years 1995 and 1998. We have analysed 
this light curve and published radial velocity data, finding that
this system is semidetached, with the secondary (less massive and less 
luminous component) filling its Roche lobe. From our analysis with the 
Wilson-Devinney code, we estimated the following masses and radii for the 
components of HV\,2543: $M_1=25.63 \pm 0.7 \, {\rm M}_{\sun}$, $R_1=15.54 \pm
0.4 \, {\rm R}_{\sun}$, $M_2=15.63 \pm 1.0 \, {\rm M}_{\sun}$ and $R_2=13.99 
\pm 0.4 \, {\rm R}_{\sun}$.
 On the basis of $B$ and $V$ 
photometry of the field stars, we found that \object{HV\,2543} is member of 
an OB association, perhaps related to which the massive binary system 
\object{Sk--67$ \degr $105} belongs.

   \keywords{binaries: eclipsing -- stars: early-type -- stars: fundamental 
parameters -- stars: individual: \object{HV 2543}  
               }  
   \end{abstract}  
  

\section{Introduction}

The analysis of light curves of eclipsing binaries, in addition  
to radial velocity data, provides fundamental knowledge about  
the masses and physical dimensions of the  stars.  Studying  massive  
binaries in the \object{Magellanic Clouds} we can learn about the evolution  
of these systems at metallicities lower than that of our galaxy.  
  
The Harvard variable \object{HV\,2543} ($\alpha=5^{\rm h}27^{\rm m}27^{\rm s}$,
$\delta=-67\degr 11\arcmin 54\arcsec$, 
J2000) is a hot binary star in the \object{LMC}. It was catalogued as an OB 
star by Sanduleak (\cite{sk}), who assigned it the identification 
--67$\degr$117. The eclipsing nature of this  
binary  was  discovered  by  Gaposhkin (\cite{gap70}),  who  published  a  
photographic  light  curve  fitting  a   period of 4.829052 days (see also  
Payne-Gaposhkin \cite {pg}).  
Photoelectric photometry was performed by Isserstedt (\cite{isser}) who  
found $V=12.92$, $(B-V)=-0.18$  and $(U-B)=-1.03$.  
Radial velocity orbit was obtained by Niemela \& Bassino (\cite{virpi}) who  
derived physical parameters of the binary components and concluded that   
HV\,2543 was a semidetached system with the less massive component 
filling its equipotential Roche surface. On the basis of their  spectroscopic 
data,  they  classified  this  system as O8V:+O9III. 
Smith Neubig \& Bruhweiler (\cite{neubig}) have published an 
UV spectral classification of LMC OB stars based on IUE data,
assigning to HV\,2543 the type O9III.
  
In this paper we present a CCD $V$ light curve for HV\,2543.  
By means  of  the combined analysis  of these data   
and previously published  radial  velocities,  we  
derive new values of the fundamental parameters of this system.  

The paper is organized as follows: in Sect. 2  we  describe  
the observations, reductions and calibration steps. In Sect. 3  
we describe the photometric results and light curve fitting. In Sect. 4 we 
discuss the results and in Sect. 5 we present our conclusions.  

  
\section{Data Acquisition and Reductions  
}  
  
\subsection {Observations}

   \begin{table}  
      \caption[]{Instrumental configuration}  
         \label{telescopio}  
\begin{flushleft}  
\begin{tabular}{ll}  
\hline  
\noalign{\smallskip}  
  
Detector                & Tektronix $1024 \times 1024$ CCD \cr  
readout noise           & $10.4 \,{\rm e}^-$  \cr  
gain                    & $7.97 \,{\rm e}^-/$ADU \cr  
filters                 & Johnson $V$ and $B$ \cr  
telescope configuration & direct CCD + focal reducer \cr  
focal relation          & 2.83 \cr  
scale                   & $0.813 \, {\rm arcsec \, pixel}^{-1}$ \cr  
field size              & diameter $\sim 9'$, circular \cr  
  
\noalign{\smallskip}  
\hline  
\end{tabular}  
\end{flushleft}  
\end{table}  

   \begin{table*}  
      \caption[]{Observing logs}  
         \label{turnos}  
  
\begin{flushleft}  
\begin{tabular}{llll}  
\hline  
\noalign{\smallskip}  
  
date  &  1995, Oct 21--25  &  1997, Nov 15--21  &  1998, Dec 2--8 \cr  
  
HJD-2450000  & 0011--0016  &  0767--0774        & 1149--1156     \cr  
number of images acquired $\times$ filter &$ 35 \times V $&$ 19 \times B + 42 \times V $&$ 36 \times V$ \cr  
range of exp. times [s] & 15--40 & 5--40 & 7--30 \cr  
  
\noalign{\smallskip}  
\hline  
\end{tabular}  
\end{flushleft}  
\end{table*}

\begin{figure}  
      \resizebox{\hsize}{!}{\includegraphics{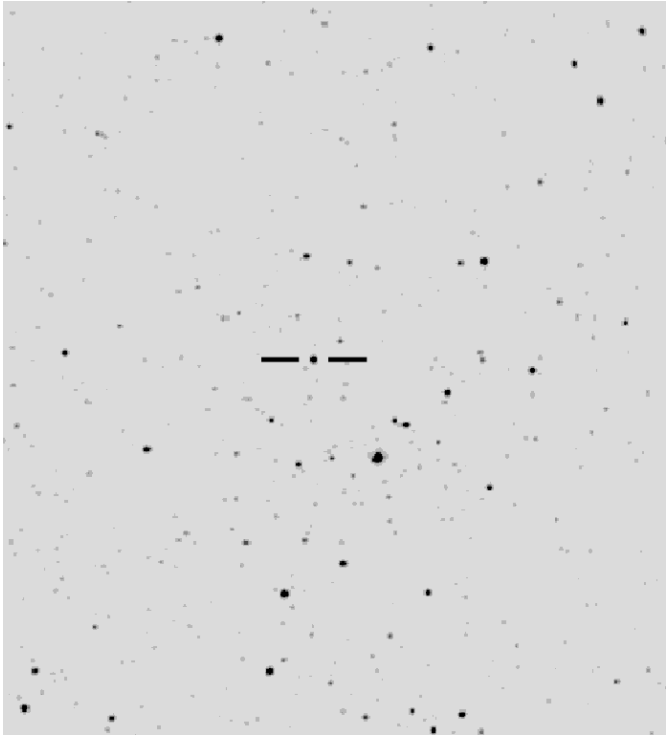}}

      \caption[]{Finding chart for HV\,2543. North is at the top and 
East is at the left. The size of the region displayed is approximately
$5.4' \times 6.4'$
              }  
         \label{chart}  
   \end{figure}

The CCD images here analysed were  acquired  with  the 2.15-m  
telescope at CASLEO, during three runs in  1995,  1997  and  1998.  
About five frames of the HV\,2543 field were obtained  each  night,  
except  during  eclipses,  when  the   observation   was   more  
intensive. During each observing night, a series of $10 \sim 15$ twilight   
flats and bias frames was also obtained. In addition, for the 1997 and 1998   
runs, $10 \sim 15$  dome flats were also obtained each night. 
Tables~\ref{telescopio} and~\ref{turnos} list details of the observations and
Fig.~\ref{chart} shows a finding 
chart for HV\,2543.  
  
\subsection{Reductions and Photometry}  
  
Reductions were performed using IRAF routines. For each  night,  
the frames were reduced in the standard  way  (overscan and bias  
corrections, flat fielding) using properly combined bias and  flat  
frames.  The  illumination  differences  between   sky-flats   and  
dome-flats were found to be less than 0.3~\% at the edges of the frames, and   
negligible at the center. Whenever high quality  
sky-flats (that  is,  zero  clouds)  were  available,  these  were  
preferred to dome-flats.  
  
For each observing run, the following steps were performed:  
  
\begin{enumerate}

\item A master image was  made  by aligning and combining  the  highest  
quality  frames (best seeing and darkest sky). On this combined  image,  
a profile fitting photometry was  performed  using  a  stand-alone  
version of DAOPHOT II (Stetson \cite{pbs87}, Stetson \cite{pbs91}).  
The resulting output was used as master  coordinate  list  in  the  
following steps.  
  
\item Aperture photometry was performed on  each  individual  frame using a
circular diaphragm of $9.5 \arcsec$ diameter.
The sky background was determined locally for each star using a sky annulus 
of radii 9.76 and $20.33 \arcsec$.
  
\item A list of instrumental magnitudes for each 
individual frame
 and for each star in the master list was constructed.  
  
\end{enumerate}  
  
   Five faint neighbour stars were detected in the combined frame within a   
radius of $9 \arcsec$ around HV\,2543, being the brightest of these some 5 
mag fainter than the binary. Light contamination from these stars slightly 
contributes to the measured magnitude of HV\,2543 (differences of the 
order of 0.014 mag). Consequently, an adequate flux correction was applied to 
the aperture photometry.

\subsection{Photometric calibration}  

No single objects were chosen as comparison  and  check  stars.  
Instead, zero point corrections were computed for each frame using  
a group  of $\sim$15  stars, 
covering widely the range of magnitudes of 
HV\,2543. 
To put all the observations in an unique instrumental system, we proceeded as
follows:

\begin{enumerate}

\item A synthetic reference system was defined using the mean instrumental 
magnitudes of the calibration stars.

\item For each frame, a zero point offset was computed and the magnitudes were
transformed to the synthetic reference system.

\item After this transformation, the RMS residuals were computed for each star.
Only the stars with the smallest residuals ($< 0.01$) were conserved as 
calibration stars.

\item We repeated these steps with the new, cleaned calibration star list.

\end{enumerate}

Since all nights were not really photometric,  this procedure allowed us to 
have a direct estimate of the internal  errors  of each  image.  

The above enumerated steps were performed independently for each observing
run. We proceeded in this way because the stars available as calibrators are 
not the same for each run, due to pointing differences. On the other hand, some
stars remain perfectly stable during a run, but show luminosity variations from
one year to another, being then inadequate as calibrators between different 
runs.

After this first stage of calibration was complete, a  list  of  
robust  estimates  of instrumental stellar  magnitudes  and their  dispersions 
was constructed for each run. After cross identification  between  the  
three lists, the most stable  stars  were  selected  to  serve  as  
calibrators between the different runs.

During the 1997 run, 19 frames through  the Johnson's $B$  band  were  also  
acquired. These frames were used to determine the transformations to the   
standard $V$ system.   

The final  transformations  to  the  standard  
system were performed by means of aperture photometry of 33 standard  stars  
-- ranging from $(B-V)=-.004$ to $2.192$ -- 
from Landolt (\cite{lan}) acquired during the photometric nights.  

We used the following transformation equations:

 $b = B + b_1 + b_2~X + b_3~(B-V)$

 $v = V + v_1 + v_2~X + v_3~(B-V)$

We adopted the mean values of the extinction coefficients given by Minniti 
et al. (\cite{minni}) and obtained the following values for the other 
coefficients:

\medskip
Nov. 19, 1997:

$b_1 = 4.2826 \pm 0.0105$

$b_3 = -0.0208 \pm 0.0115$

$v_1 = 3.4446 \pm 0.0051$

$v_3 = -0.0456 \pm 0.0055$
\medskip

Nov. 20, 1997:

$b_1 = 4.2986  \pm 0.0076 $

$b_3 = -0.03116 \pm 0.0086 $

$v_1 = 3.4502 \pm 0.0057 $

$v_3 = -0.0535 \pm 0.0067 $
\medskip

The RMS residuals of the $V$ transformations were 0.017 mag. We first shifted 
our unified instrumental system to that corresponding to a $V$ frame acquired 
immediately before a series of B exposures, on Nov. 19, 1997. Then, we applied 
the standard transformations derived for that night.
The Nov. 20 transformations were used only to check the quality of the first 
ones. 

Our photometry of HV\,2543 is presented in Table~\ref{tlc}. In successive 
columns the heliocentric julian day, standard V magnitude, internal
errors, seeing and airmass are given. The errors ($\sigma_{\rm i}$) account for
the internal photometric errors estimated by DAOPHOT and those of the
transformation to an unified instrumental system. They do not include the error
of the transformation to the standard system.


   \begin{table*}  
      \caption[]{$V$ light photometry of HV\,2543.
}  
         \label{tlc}  
  
\begin{flushleft}  
\begin{tabular}{rrrrlrrrrl}  
\hline  
\noalign{\smallskip}  
  
 HJD       &$    V    $&$ \sigma_{\rm i}$&   fwhm    &$  X    $&
 HJD       &$    V    $&$\sigma_{\rm i}$&   fwhm    &$  X    $\cr  
$  2450000+ $&           &               &$  \arcsec  $&               &
$  2450000+ $&           &               &$  \arcsec  $&               \cr
\noalign{\smallskip}  
\hline  
\noalign{\smallskip}  
  
  
$    11.680 $&$  12.807 $&$   0.009 $&$   2.98 $&$   1.51 $&$   771.812 $&$  12.919 $&$   0.005 $&$   2.56 $&$   1.26 $\cr
$    11.726 $&$  12.803 $&$   0.013 $&$   2.34 $&$   1.36 $&$   771.850 $&$  12.905 $&$   0.006 $&$   2.61 $&$   1.31 $\cr
$    11.768 $&$  12.806 $&$   0.013 $&$   2.44 $&$   1.26 $&$   772.637 $&$  12.872 $&$   0.006 $&$   2.66 $&$   1.37 $\cr
$    11.812 $&$  12.790 $&$   0.012 $&$   3.31 $&$   1.23 $&$   772.681 $&$  12.872 $&$   0.008 $&$   2.84 $&$   1.28 $\cr
$    12.668 $&$  12.988 $&$   0.010 $&$   3.03 $&$   1.54 $&$   772.753 $&$  12.881 $&$   0.006 $&$   2.61 $&$   1.23 $\cr
$    12.730 $&$  13.032 $&$   0.010 $&$   3.98 $&$   1.34 $&$   772.806 $&$  12.884 $&$   0.006 $&$   2.78 $&$   1.25 $\cr
$    12.768 $&$  13.074 $&$   0.009 $&$   3.94 $&$   1.27 $&$   772.856 $&$  12.887 $&$   0.007 $&$   3.04 $&$   1.33 $\cr
$    12.810 $&$  13.119 $&$   0.010 $&$   4.15 $&$   1.23 $&$   773.593 $&$  13.759 $&$   0.006 $&$   2.43 $&$   1.51 $\cr
$    12.838 $&$  13.156 $&$   0.013 $&$   4.67 $&$   1.23 $&$   773.658 $&$  13.763 $&$   0.007 $&$   2.00 $&$   1.32 $\cr
$    12.864 $&$  13.204 $&$   0.016 $&$   4.71 $&$   1.23 $&$   773.708 $&$  13.674 $&$   0.004 $&$   2.41 $&$   1.25 $\cr
$    13.683 $&$  12.912 $&$   0.010 $&$   2.88 $&$   1.47 $&$   773.729 $&$  13.629 $&$   0.004 $&$   2.51 $&$   1.23 $\cr
$    13.732 $&$  12.900 $&$   0.010 $&$   3.30 $&$   1.33 $&$   773.742 $&$  13.609 $&$   0.005 $&$   2.42 $&$   1.23 $\cr
$    13.773 $&$  12.883 $&$   0.013 $&$   3.76 $&$   1.26 $&$   773.754 $&$  13.572 $&$   0.005 $&$   3.24 $&$   1.23 $\cr
$    13.852 $&$  12.865 $&$   0.009 $&$   3.14 $&$   1.23 $&$   773.770 $&$  13.540 $&$   0.006 $&$   2.61 $&$   1.23 $\cr
$    14.665 $&$  12.879 $&$   0.006 $&$   3.38 $&$   1.53 $&$   773.780 $&$  13.506 $&$   0.005 $&$   2.69 $&$   1.23 $\cr
$    14.718 $&$  12.894 $&$   0.006 $&$   4.07 $&$   1.36 $&$   773.791 $&$  13.479 $&$   0.007 $&$   3.11 $&$   1.24 $\cr
$    14.755 $&$  12.903 $&$   0.005 $&$   3.76 $&$   1.28 $&$   773.804 $&$  13.457 $&$   0.008 $&$   3.57 $&$   1.25 $\cr
$    14.791 $&$  12.919 $&$   0.007 $&$   3.94 $&$   1.24 $&$   773.816 $&$  13.414 $&$   0.005 $&$   2.63 $&$   1.27 $\cr
$    14.861 $&$  12.940 $&$   0.015 $&$   5.19 $&$   1.24 $&$   773.833 $&$  13.380 $&$   0.007 $&$   2.60 $&$   1.29 $\cr
$    15.577 $&$  13.607 $&$   0.014 $&$   4.04 $&$   2.07 $&$   773.853 $&$  13.345 $&$   0.011 $&$   3.46 $&$   1.33 $\cr
$    15.584 $&$  13.588 $&$   0.011 $&$   3.24 $&$   2.01 $&$  1149.809 $&$  12.996 $&$   0.016 $&$   2.19 $&$   1.30 $\cr
$    15.590 $&$  13.570 $&$   0.011 $&$   3.08 $&$   1.96 $&$  1149.837 $&$  13.020 $&$   0.005 $&$   2.36 $&$   1.36 $\cr
$    15.595 $&$  13.557 $&$   0.010 $&$   3.10 $&$   1.92 $&$  1150.848 $&$  12.922 $&$   0.012 $&$   2.09 $&$   1.40 $\cr
$    15.603 $&$  13.535 $&$   0.009 $&$   2.75 $&$   1.86 $&$  1151.616 $&$  12.810 $&$   0.007 $&$   2.27 $&$   1.34 $\cr
$    15.609 $&$  13.523 $&$   0.010 $&$   3.16 $&$   1.82 $&$  1151.670 $&$  12.814 $&$   0.005 $&$   2.23 $&$   1.25 $\cr
$    15.618 $&$  13.496 $&$   0.009 $&$   3.67 $&$   1.76 $&$  1151.730 $&$  12.820 $&$   0.004 $&$   2.05 $&$   1.23 $\cr
$    15.628 $&$  13.462 $&$   0.010 $&$   3.48 $&$   1.70 $&$  1151.785 $&$  12.825 $&$   0.007 $&$   2.48 $&$   1.27 $\cr
$    15.643 $&$  13.432 $&$   0.009 $&$   3.32 $&$   1.62 $&$  1151.853 $&$  12.838 $&$   0.007 $&$   2.20 $&$   1.42 $\cr
$    15.675 $&$  13.361 $&$   0.010 $&$   2.34 $&$   1.48 $&$  1152.644 $&$  13.386 $&$   0.006 $&$   2.57 $&$   1.28 $\cr
$    15.693 $&$  13.305 $&$   0.011 $&$   2.79 $&$   1.42 $&$  1152.662 $&$  13.414 $&$   0.006 $&$   2.54 $&$   1.25 $\cr
$    15.717 $&$  13.265 $&$   0.010 $&$   3.20 $&$   1.35 $&$  1152.690 $&$  13.443 $&$   0.006 $&$   1.93 $&$   1.23 $\cr
$    15.738 $&$  13.220 $&$   0.012 $&$   3.81 $&$   1.31 $&$  1152.708 $&$  13.454 $&$   0.005 $&$   2.18 $&$   1.23 $\cr
$    15.787 $&$  13.151 $&$   0.010 $&$   3.73 $&$   1.24 $&$  1152.728 $&$  13.457 $&$   0.004 $&$   2.42 $&$   1.23 $\cr
$    15.834 $&$  13.077 $&$   0.011 $&$   4.11 $&$   1.23 $&$  1152.737 $&$  13.456 $&$   0.005 $&$   2.19 $&$   1.23 $\cr
$    15.865 $&$  13.048 $&$   0.010 $&$   3.32 $&$   1.24 $&$  1152.750 $&$  13.447 $&$   0.006 $&$   2.10 $&$   1.24 $\cr
$   767.694 $&$  12.848 $&$   0.009 $&$   2.35 $&$   1.28 $&$  1152.761 $&$  13.434 $&$   0.005 $&$   1.97 $&$   1.25 $\cr
$   767.768 $&$  12.857 $&$   0.006 $&$   3.05 $&$   1.23 $&$  1152.786 $&$  13.401 $&$   0.004 $&$   2.21 $&$   1.28 $\cr
$   767.808 $&$  12.864 $&$   0.008 $&$   3.68 $&$   1.24 $&$  1152.799 $&$  13.384 $&$   0.007 $&$   2.14 $&$   1.30 $\cr
$   767.849 $&$  12.866 $&$   0.005 $&$   3.09 $&$   1.29 $&$  1152.812 $&$  13.358 $&$   0.003 $&$   1.84 $&$   1.32 $\cr
$   768.627 $&$  13.451 $&$   0.009 $&$   3.11 $&$   1.44 $&$  1152.829 $&$  13.335 $&$   0.004 $&$   2.01 $&$   1.36 $\cr
$   768.685 $&$  13.607 $&$   0.007 $&$   2.77 $&$   1.29 $&$  1152.855 $&$  13.295 $&$   0.008 $&$   2.35 $&$   1.44 $\cr
$   768.748 $&$  13.743 $&$   0.007 $&$   2.32 $&$   1.23 $&$  1153.687 $&$  12.842 $&$   0.008 $&$   1.57 $&$   1.23 $\cr
$   768.803 $&$  13.790 $&$   0.006 $&$   2.38 $&$   1.24 $&$  1153.733 $&$  12.833 $&$   0.005 $&$   1.79 $&$   1.23 $\cr
$   768.825 $&$  13.786 $&$   0.006 $&$   2.30 $&$   1.26 $&$  1153.766 $&$  12.832 $&$   0.007 $&$   2.05 $&$   1.25 $\cr
$   768.840 $&$  13.758 $&$   0.006 $&$   2.32 $&$   1.28 $&$  1153.821 $&$  12.828 $&$   0.006 $&$   2.24 $&$   1.35 $\cr
$   768.856 $&$  13.738 $&$   0.008 $&$   2.17 $&$   1.31 $&$  1153.857 $&$  12.825 $&$   0.006 $&$   2.86 $&$   1.45 $\cr
$   768.868 $&$  13.718 $&$   0.015 $&$   2.19 $&$   1.33 $&$  1154.704 $&$  13.044 $&$   0.007 $&$   2.15 $&$   1.23 $\cr
$   769.639 $&$  12.871 $&$   0.007 $&$   2.93 $&$   1.39 $&$  1154.738 $&$  13.082 $&$   0.007 $&$   2.80 $&$   1.23 $\cr
$   769.716 $&$  12.860 $&$   0.015 $&$   3.01 $&$   1.25 $&$  1154.797 $&$  13.151 $&$   0.009 $&$   2.50 $&$   1.30 $\cr
$   769.792 $&$  12.847 $&$   0.011 $&$   3.37 $&$   1.23 $&$  1154.843 $&$  13.228 $&$   0.005 $&$   2.36 $&$   1.42 $\cr
$   769.835 $&$  12.836 $&$   0.006 $&$   2.86 $&$   1.28 $&$  1155.656 $&$  12.962 $&$   0.007 $&$   2.16 $&$   1.25 $\cr
$   770.718 $&$  12.937 $&$   0.009 $&$   1.96 $&$   1.24 $&$  1155.710 $&$  12.935 $&$   0.005 $&$   1.94 $&$   1.23 $\cr
$   770.766 $&$  12.963 $&$   0.006 $&$   2.39 $&$   1.23 $&$  1155.746 $&$  12.923 $&$   0.006 $&$   1.86 $&$   1.24 $\cr
$   770.814 $&$  12.990 $&$   0.005 $&$   2.21 $&$   1.25 $&$  1155.774 $&$  12.907 $&$   0.006 $&$   1.82 $&$   1.27 $\cr
$   770.845 $&$  13.020 $&$   0.005 $&$   1.78 $&$   1.30 $&$  1155.823 $&$  12.886 $&$   0.007 $&$   2.64 $&$   1.37 $\cr
$   771.701 $&$  12.951 $&$   0.012 $&$   2.27 $&$   1.26 $&$  1155.849 $&$  12.883 $&$   0.005 $&$   2.66 $&$   1.44 $\cr
$   771.777 $&$  12.923 $&$   0.011 $&$   2.47 $&$   1.23 $\cr

\noalign{\smallskip}  
\hline  
\end{tabular}  
\end{flushleft}  
\end{table*}  
  

\section {Results}  
  
\subsection{Ephemeris}
  
   We have observed one primary minimum during 1997, at   
HJD$=2\,450\,768 \fd 802 \pm 0.001$. Combining this value with that given by   
Payne-Gaposchkin (\cite{pg}) we obtained  a new ephemerides for HV\,2543:  
\medskip  
  
$E_0 = 2 \, 450 \, 768 \fd 802 \pm 0.001$  
  
$P=4 \fd 829 \, 046 \pm 0.000\,004$

\medskip  
We have used this ephemeris in calculating the phases for  the radial velocity 
and photometric data.

\subsection {Light and radial velocity curve analysis}

  \begin{figure*}  
      \resizebox{\hsize}{!}{\includegraphics{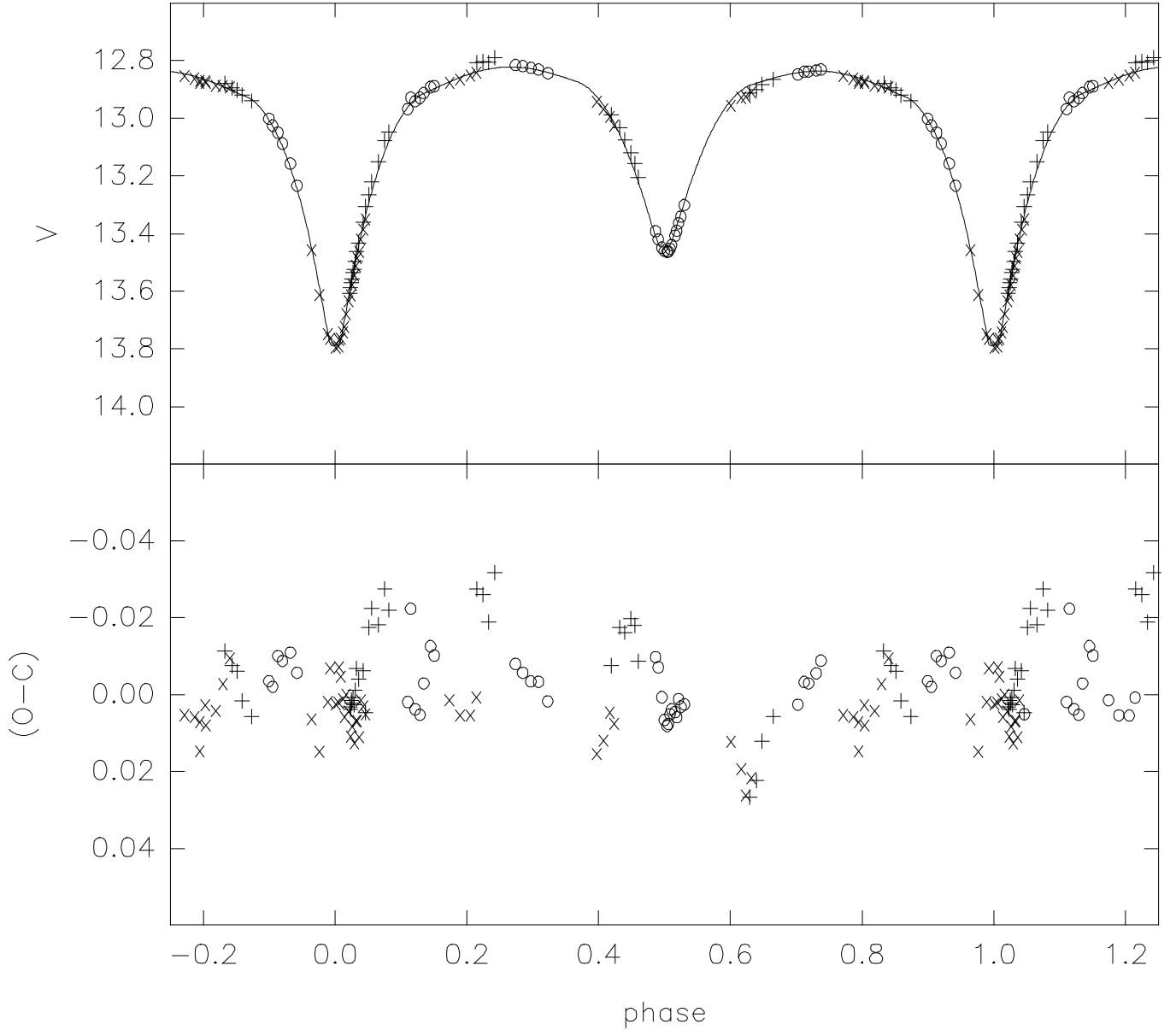}}
      \caption[]{   
Top: Observed and modelled light curve for HV 2543.   
Bottom: (O-C) residuals for the light curve.  
The symbols $+$, $\times$ and {\large $\circ$} correspond to the data collected
during 1995, 1997 and 1998 respectively.  
}  
         \label{lc}
   \end{figure*}  

   \begin{table}  
      \caption[]{Model Parameters}  
         \label{modelo}  
\begin{flushleft}  
\begin{tabular}{ll}  
\hline  
\noalign{\smallskip}  
$a$                    &$ 41.5 \pm 0.5 \, {\rm R}_{\sun}  $ \cr  
$V_{\gamma}$           &$ 293.2 \pm 6 \, {\rm km s}^{-1}    $ \cr  
$i$                    &$ 89 \pm 1 \degr      $ \cr  
$q~(M_2/M_1)$          &$ 0.61 \pm 0.05    $ \cr  
\noalign{\medskip}  
$T_1$                  &$ 36000 \sim 34700 \,{\rm K}$ ~(adopted) \cr  
$\Omega_1$             &$ 3.363 \pm 0.06   $ \cr  
$g_1$                  & 1.00 ~(adopted)   \cr  
$A_1$                  & 1.00 ~(adopted)   \cr  
$x_1~^{\rm b}$               & 0.541~ (adopted)  \cr  
\noalign{\medskip}  
$T_2$                  &$ 29200 \sim 28180 \, {\rm K} ~^{\rm a}$ \cr  
$\Omega_2$             &$ 3.082 \pm 0.12   $ \cr  
$g_2$                  & 1.00 ~(adopted)   \cr  
$A_2$                  & 1.00 ~(adopted)   \cr  
$x_2~^{\rm b}$               & 0.561~ (adopted)  \cr  
\noalign{\medskip}  
\hline  
\noalign{\smallskip}  
\noalign{One hot spot solution (in star 1):}  
\noalign{\smallskip}  
colatitude &$ 90 \degr $ \cr  
longitude &$ 253.4 \degr $ \cr  
angular~ radius &$ 35.8\degr $ \cr  
temp.~ factor &$ 1.053 $ \cr  
\noalign{\medskip}  
\noalign{Two hot spot solution:}  
\noalign{\smallskip}  
\noalign{For star 1:}  
\noalign{\smallskip}  
colatitude &$ 90 \degr $ \cr  
longitude &$ 270 \degr $ \cr  
angular~ radius &$ 39.1 \degr $ \cr  
temp.~ factor & 1.032  \cr  
  
\noalign{\medskip}  
\noalign{For star 2:}  
\noalign{\smallskip}  
  
colatitude &$ 90 \degr $ \cr  
longitude &$ 0 \degr $ \cr  
angular~ radius &$ 80 \degr $ \cr  
temp.~ factor & 1.030  \cr  
  
\noalign{\smallskip}  
\hline  
\end{tabular}  
\begin{list}{}{}  
\item[$^{\rm a}$] Values for $T_2$ resulting for the above quoted values adopted for 
$T_1$ respectively.  
\item[$^{\rm b}$]~$x_1$ and $x_2$ stand for the bolometric limb-darkening   
coefficients.  
\end{list}    
\end{flushleft}  
\end{table}

The $V$ light curve of HV\,2543 is typical of near-contact binaries,  
with different depths of the minima and continuous variations of  
light in out-of-eclipse portions. Also noticeable is the O'Connell effect,   
i.e., the small difference in magnitude between both maxima (see Davidge  
\& Milone \cite{damilo}). We solved   
simultaneously the light and radial velocity curves using the   
Wilson \& Devinney code (hereafter WD), which is very well suited for the study
of close binary systems (Wilson \& Devinney \cite{wd}, Wilson \cite{wil}). 
Unit weight was given to photometric CCD observations since 
they all were collected with the same instrumental configuration and have 
comparable quality.
A unit weight was also given to most of the
radial velocity points except for those 
near eclipses or 
indicated by Niemela 
\& Bassino (\cite{virpi}) as less confident. We also corrected the data points 
corresponding to the phase $\phi=0.05$, that are inverted (i.e., the O8   
velocity corresponds to the O9 and vice-versa) in Niemela \& Bassino   
(\cite{virpi}). The relation between photometric and   
spectroscopic weights was given through the values of the mean standard  
deviations (sigma) of the data which were estimated to be 17 and 12 
km s$^{-1}$ for   
the radial velocities of the primary and secondary components, respectively,  
and 0.02 mag for the light points. Only the ratios between these values are  
significant.  
  
  The radii that result from our preliminary light curve analysis indicate that
the O8 component is more alike an O8III star. Chlebowski \& Garmany 
(\cite{chg}) give a  
temperature of 36000\,K for an O8III star. Given that 
Schmidt-Kaler (1982) gives a lower temperature for such spectral type, we also 
performed an analysis assigning a 34700\,K temperature for the   
primary.  
Standard values of bolometric albedos, $A = 1.0$ (Rucinski \cite{ruc}), and   
gravity darkening coefficients, $g = 1.0$ (Lucy \cite{luc}), for radiative   
envelopes were used. Linear limb-darkening coefficients were determined from   
tables by van Hamme (\cite{ham}). These parameters were not adjusted. The   
adjustable parameters in our computations were: $a$ (semimajor axis), 
$V_{\gamma}$ (systemic radial velocity), $i$ (orbital inclination),  
$T_2$ (temperature of the secondary component), $q$ (mass-ratio), $\Omega_1$   
and $\Omega_2$ (modified potential of both components) and $L_1$ (luminosity 
of the primary). The first approaches to fit the observations were made by 
means of the light curve (LC) program until an acceptable fit to the $V$ light 
curve was obtained.  

     Then we proceeded with the differential corrections (DC) code.
We first adjusted the parameters $a$, $V_{\gamma}$ and $q$ using only the 
out-of-eclipses radial velocity data. Thereupon, we left those parameters 
fixed and fitted $i$, $T_2$, $\Omega_1$, $\Omega_2$ and $L_1$, using only the
light curve. Then we re-computed the first set of parameters leaving fixed
the second set, and iterated this procedure until the corrections were smaller 
than their errors. A better solution was achieved allowing the fit of $q$ with
the photometric data, that is, including $q$ within the second parameter set.
Mode 2 (Leung \& Wilson \cite{lw})   
which stands for detached systems was used at the beginning, but after a few   
runs of the DC program the computations clearly evolved towards a semidetached
configuration with the secondary component filling its critical Roche lobe.  
Thus, mode 5 of the WD code was employed until the final solutions were  
found. In this mode, the parameter $\Omega_2$ cannot be adjusted and it is set 
equal to the critical filling-lobe value. The solutions never converged to  
contact configurations.  
  
\begin{figure}  
      \resizebox{\hsize}{!}{\includegraphics{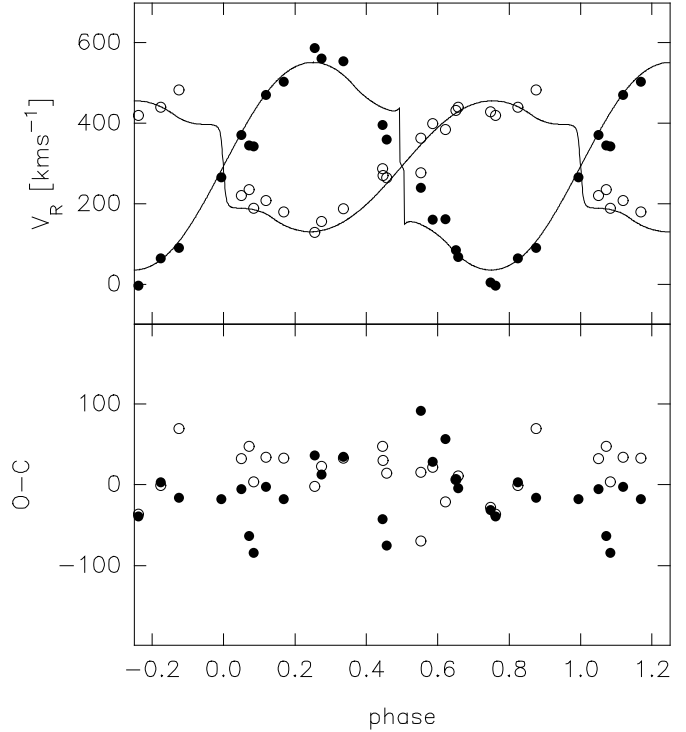}}
      \caption[]{Top: Observed and modelled radial velocity curve for   
HV\,2543. Bottom: (O-C) residuals for the radial velocities. 
Hollow circles correspond to the primary component and filled ones stand for
the secondary.
}  
         \label{vc}  
   \end{figure}

 Since some systematic  
differences were obtained between the observed and modelled light curves  
(due to the O'Connell effect), we tried to improve the binary modelling by  
including one or two hot spots on the surface of the stars. The spots   
were placed at the equator of the stars (colatitude\,$ = 90\degr$) while   
the other spot parameters where adjusted with the WD code.   
 The range of the adjusted temperature for the secondary star, 
$T_2$, depends on the adopted value for $T_1$. To estimate the errors of the   
other parameters we considered the differences between the values that arised  
from different solutions 
(i.e., using the ``spectroscopic'' value of $q$, 0.55,
or leaving it to increase until 0.64, to allow a better fit to the photometric
data). 
The adopted solution, with $q=0.61$, is a compromise between the optimal fit
of the light curve and the tolerable deviation from the spectroscopic data.
Table~\ref{modelo}, lists the model parameters.  
Fig.~\ref{lc} and ~\ref{vc} show the modelled light and radial velocity  
curves, respectively, derived from the 
one spot solution 
and plotted along  
with the observations, and their corresponding residuals (O-C).  
  
 The final value of the orbital inclination is incidentally   
high, very close to $90\degr$. This fact should produce a total eclipse at 
the secondary minimum, but, as the dimensions of both components are quite   
similar, this is not a noticeable feature of the observed light curve.

 Absolute values of the masses, dimensions and bolometric magnitudes of 
HV\,2543 were computed from the 
one spot 
solution. They are listed in 
Table~\ref{size} where $R_1$ and $R_2$ correspond to mean values of the derived
polar, back and side radii of the stars. The final configuration of HV\,2543 
shows two components of similar dimensions but different masses and 
temperatures, with the secondary less massive star filling its Roche lobe.

     \begin{table}  
      \caption[]{Star dimensions}  
         \label{size}  
\begin{flushleft}  
\begin{tabular}{ll}  
\hline  
\noalign{\smallskip}  
  
$M_1       $&$  25.63 \pm 0.7~{\rm M}_{\sun}  $\cr   
$R_1       $&$  15.54 \pm 0.4~{\rm R}_{\sun}  $\cr  
$M_{bol~1} $&$ -9.12 \sim -8.90~^{\rm a}      $\cr        
$\log g_1 [cgs]  $&$ 3.46 \pm 0.04        $\cr  
\noalign{\smallskip}  
$M_2   $&$  15.63 \pm 1.0~ {\rm M}_{\sun}  $\cr   
$ R_2  $&$  13.99 \pm 0.4 {\rm R}_{\sun}   $\cr  
$ M_{bol~2} $&$ -7.98 \sim -7.82  ~^{\rm a}   $\cr        
$\log g_2 [cgs]   $&$ 3.34 \pm 0.01   $\cr  
  
\noalign{\medskip}  
\hline  
\end{tabular}  
\begin{list}{}{}  
\item[$^{\rm a}$] The range in the derived bolometric magnitudes corresponds to
the range in the values adopted for $T_1$.  
\end{list}    
\end{flushleft}  
\end{table}

\subsection {$BV$ Field photometry}  
\begin{figure}  
      \resizebox{\hsize}{!}{\includegraphics{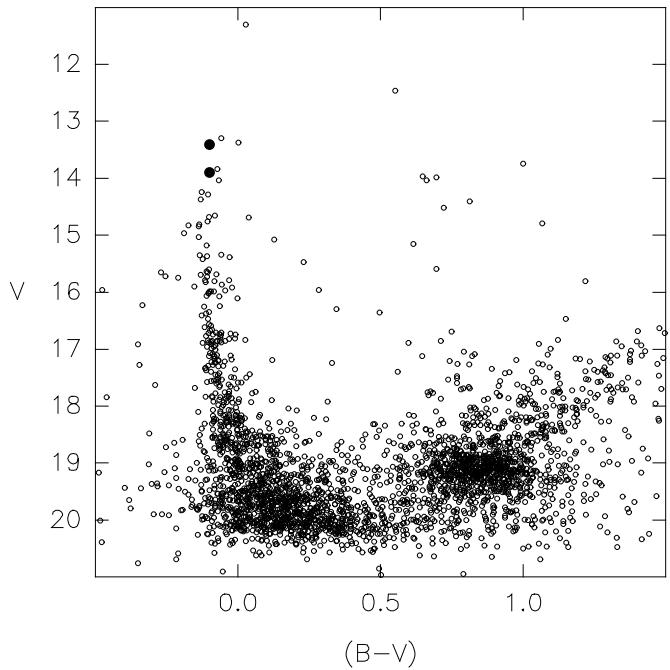}}

      \caption[]{Color-Magnitude diagram for the field surrounding HV\,2543.   
The two filled circles represent the magnitude estimates for the components  
of HV\,2543. Magnitudes and colours are not corrected for extinction.   
              }  
         \label{colomag}  
   \end{figure}

We combined the $B$ frames acquired during 1997 to make a master $B$ image, 
on which we performed a profile fitting photometry. We used this photometry 
together  with that performed on the $V$ master frame to generate a 
colour-magnitude diagram, which is shown in Fig.~\ref{colomag}.

In this figure, the presence of an OB association covering the whole   
frame ($9 \arcmin$ of diameter) is evident. 
%
%
Ostrov et al. (\cite{yoetal}) found evidence of a stellar association 
surrounding the massive binary system \object{Sk--67$\degr$105}. 
The relatively small angular distance between the two stellar groups 
($\sim 9.5 \arcmin$)  suggests that they are probably related.

We have obtained spectroscopic data for five stars in the neighbourhood of~ 
Sk-67$\degr$105. For these stars, we have estimated the reddening by 
comparison between the intrinsic $(B-V)_0$ colours given by 
Schmidt-Kaler (\cite{s-k}) and FitzGerald (\cite{fg}) and the observed colours.
We derived $E(B-V)=0.17\pm0.015$, while for HV\,2543 itself we obtained 
$E(B-V)=0.20$. A detailed study of these OB associations will be presented
in a forthcoming paper.

\section{Discussion}
  
   We note that the radius derived for the O8 component is substantially larger
than the one obtained by Niemela \& Bassino (\cite{virpi}). In fact, the   
spectroscopic data suggest that the luminosity of the O9 
star is larger than that of the O8 star, which caused Niemela \& 
Bassino to refer to the O9 component as ``primary'', even though it is the
one presenting the largest radial velocity amplitude, and consequently
the less massive component of the binary system.
Hence, we explored alternative solutions to the light curve
that could yield a smaller radius for the O8 component. 
The values  $R_1 \sim 12.75$, $R_2 \sim 15.31$, $M_1 \sim 21.91$ and 
$M_2 \sim 19.34$, in solar units, provide a good fit to the light curve, 
although they require a value of $q$ near 0.9, which is not compatible with 
the radial velocity data. Hence we have discarded this solution, which on the
other hand, implies a rather low value  of the distance modulus, namely 18.18, 
assuming $A_V = 0.62$ (see below) and adopting the temperature scale of
Chlebowski \& Garmany (\cite{chg}). This value would result even smaller if
a lower temperature scale is adopted.

  On the other hand, the temperature difference between the two components of 
HV\,2543 resulting from the light curve analysis is larger than that suggested 
by the corresponding spectral types. This fact might be due to the difference 
between the spectral features  of the unperturbed back sides and the 
heated inner sides of the stars.
   Also should we have in mind that
the mass transfer history of the system might account for significant
departures of the He abundances relative to those regarded as normal
for LMC members, and this effect could 
influence some spectral lines of the secondary component.



To solve these puzzles, high dispersion spectroscopy would be desirable.  
 
%

The semimajor axis and star dimensions of this system are alike those   
determined by Pritchard et al. (\cite{prit}) for \object{HV\,2241}, but the 
masses are   
somewhat smaller. We presume that these systems have experienced case A mass  
transfer, being now in the slow stage of mass exchange. In these cases, the  
mass gainer should be indistinguishable of a normal star, excepting that   
it would stand on an isochrone corresponding to a shorter age  (see Vanbeveren 
et al. \cite{vbv}).  
From the stellar models of Schaerer et al. (\cite{sch}) (for single stars) we 
found that a $25 {\rm M}_{\sun}$ star takes $\sim 6.7$ Myr until its radius 
grows 
to $15{\rm R}_{\sun}$, but such star would have an effective temperature of 
only 30000\,K. The radius and effective temperature derived for the O8
star are consistent whit those of a single star of some 3.7 Myrs and   
$\sim 40 {\rm M}_{\sun}$, completely out of the range of masses compatible 
with the 
radial velocity data.  We note that a similar problem arises from the 
analysis of \object{AB Crucis} (Lorenz et al. \cite{lor}).

It is clear that there are a series of phenomena that we do not fully 
understand, and consequently, our analysis can not be considered definitive. 
The errors given for the derived parameters must be considered with caution. 
True errors are not easy to 
estimate analytically, since they depend on the importance of phenomena that
are not properly accounted for, such as wind shocks, uncertainty of the 
adopted temperature scale, radiation pressure effects, etc.
In fact, the sizes and positions of the spots used to model empirically
the O'Connell effect are rather arbitrary, and an equally satisfactory 
solution could   
be found with other parameters, but these details do not affect meaningly 
the derived star dimensions.

From our photometry we determine $(B-V)=-0.11 \pm 0.015$ for HV\,2543. 
This value is somewhat redder than $-0.18$, obtained by Isserstedt 
(\cite {isser}). However, given that Isserstedt does not detect the variability
of HV\,2543, we think that both measures are still in reasonable agreement.
If we assume a $(B-V)_0$ of -0.31 (Schmidt-Kaler \cite{s-k}), $R=3.1$   
(Koornneef \cite{koo}) then it results an $A_V=0.62$ for HV\,2543.   
This value does not depend on which temperature scale we  adopt, since for 
the range of temperatures of the O-type  stars the $(B-V)$ colours are 
degenerated.

Estimating the bolometric corrections according to 
Massey \& Hunter (\cite{mh}), we derive a distance modulus of 
$(m-M)_0=18.31 \pm 0.2$ to $18.40 \pm 0.2$, depending on the adopted 
temperature scale. The error accounts for the uncertainties in the estimates 
of $R$ and the bolometric corrections.
This distance modulus must be considered with
caution, since this system has experienced strong mass transfer and exhibits 
the above mentioned anomalies.

\section {Conclusions}  
  
We have presented a CCD $V$  light  curve  of  the   
eclipsing binary HV\,2543 (Sk-67$\degr$117) in the LMC. The light curve   
appearance is almost symmetric, with a slight ($\sim 0.02~ V$ magnitudes)  
O'Connell effect. From $BV$ photometry of the surrounding field, we found that 
HV\,2543 probably belongs to an OB association not previously identified. 

We have analysed this light curve and the published radial velocity data using 
the Wilson-Devinney code, finding that HV\,2543 is a  semidetached  system  
with the less massive and less luminous member filling  its  Roche  
lobe. The primary minimum occurs when the O8 component is behind  
the O9.
From our  analysis,  we  derived  fundamental  
parameters for this system, obtaining $M_1=25.63 \pm 0.7$, $M_2=15.63 \pm 1.0$,
$R_1=15.54 \pm 0.4$ and $R_2=13.99 \pm 0.4$ in solar units.
Some discrepancies between the present results and those derived from the 
previous spectroscopic analysis (Niemela \& Bassino \cite{virpi}), should be 
thoroughly addressed in future works.

We  found  that this system has experienced significant mass exchange, being   
the present most massive star the originally less massive.

\begin{acknowledgements}  

The authors acknowledge use at CASLEO of the CCD and data acquisition system   
supported under U.S. National Science Foundation grant AST-90-15827 to   
R. M. Rich. The focal reducer in use at CASLEO  was kindly provided by   
Dr. M. Shara. This research has made use of the Astronomical Data Center 
catalogs. We are indebted to the staff of CASLEO for valuable help during 
the observing runs. We are grateful to V.S. Niemela for her vigorous comments 
made during the preparation of this paper. We would also like to thank to the 
referee for the suggestions that allowed to improve the presentation of
the paper.

\end{acknowledgements}

  
\end{document}